\RequirePackage{lineno}
\documentclass[aps,prl,twocolumn,showpacs,superscriptaddress,groupedaddress]{revtex4}  
\usepackage{graphicx}  
\usepackage{dcolumn}   
\usepackage{bm}        
\usepackage{amssymb}   
\usepackage{multirow}
\usepackage{epsfig}
\usepackage{amsmath}
\usepackage[colorlinks, linkcolor=blue]{hyperref}

\newcommand{\DY}{p\bar{p}\rightarrow Z/\gamma^{*}\rightarrow \mu^{+}\mu^{-}}
\newcommand{\effstw}{\ensuremath{\sin^2\theta_{\text{eff}}^{\text{$\ell$}}}}

\newcommand{\stw}{\ensuremath{\sin^2 \theta_{W}}}

\begin{document}

\lefthyphenmin=2
\righthyphenmin=2

\widetext

\hspace{5.2in} \mbox{FERMILAB-PUB-17-434-E}
\title{Measurement of the Effective Weak Mixing Angle in
$\boldsymbol{p\bar{p}\rightarrow Z/\gamma^{*}\rightarrow \ell^{+} \ell^{-}}$ Events }

%
\affiliation{LAFEX, Centro Brasileiro de Pesquisas F\'{i}sicas, Rio de Janeiro, RJ 22290, Brazil}
\affiliation{Universidade do Estado do Rio de Janeiro, Rio de Janeiro, RJ 20550, Brazil}
\affiliation{Universidade Federal do ABC, Santo Andr\'e, SP 09210, Brazil}
\affiliation{University of Science and Technology of China, Hefei 230026, People's Republic of China}
\affiliation{Universidad de los Andes, Bogot\'a, 111711, Colombia}
\affiliation{Charles University, Faculty of Mathematics and Physics, Center for Particle Physics, 116 36 Prague 1, Czech Republic}
\affiliation{Czech Technical University in Prague, 116 36 Prague 6, Czech Republic}
\affiliation{Institute of Physics, Academy of Sciences of the Czech Republic, 182 21 Prague, Czech Republic}
\affiliation{Universidad San Francisco de Quito, Quito 170157, Ecuador}
\affiliation{LPC, Universit\'e Blaise Pascal, CNRS/IN2P3, Clermont, F-63178 Aubi\`ere Cedex, France}
\affiliation{LPSC, Universit\'e Joseph Fourier Grenoble 1, CNRS/IN2P3, Institut National Polytechnique de Grenoble, F-38026 Grenoble Cedex, France}
\affiliation{CPPM, Aix-Marseille Universit\'e, CNRS/IN2P3, F-13288 Marseille Cedex 09, France}
\affiliation{LAL, Univ. Paris-Sud, CNRS/IN2P3, Universit\'e Paris-Saclay, F-91898 Orsay Cedex, France}
\affiliation{LPNHE, Universit\'es Paris VI and VII, CNRS/IN2P3, F-75005 Paris, France}
\affiliation{CEA Saclay, Irfu, SPP, F-91191 Gif-Sur-Yvette Cedex, France}
\affiliation{IPHC, Universit\'e de Strasbourg, CNRS/IN2P3, F-67037 Strasbourg, France}
\affiliation{IPNL, Universit\'e Lyon 1, CNRS/IN2P3, F-69622 Villeurbanne Cedex, France and Universit\'e de Lyon, F-69361 Lyon CEDEX 07, France}
\affiliation{III. Physikalisches Institut A, RWTH Aachen University, 52056 Aachen, Germany}
\affiliation{Physikalisches Institut, Universit\"at Freiburg, 79085 Freiburg, Germany}
\affiliation{II. Physikalisches Institut, Georg-August-Universit\"at G\"ottingen, 37073 G\"ottingen, Germany}
\affiliation{Institut f\"ur Physik, Universit\"at Mainz, 55099 Mainz, Germany}
\affiliation{Ludwig-Maximilians-Universit\"at M\"unchen, 80539 M\"unchen, Germany}
\affiliation{Panjab University, Chandigarh 160014, India}
\affiliation{Delhi University, Delhi-110 007, India}
\affiliation{Tata Institute of Fundamental Research, Mumbai-400 005, India}
\affiliation{University College Dublin, Dublin 4, Ireland}
\affiliation{Korea Detector Laboratory, Korea University, Seoul, 02841, Korea}
\affiliation{CINVESTAV, Mexico City 07360, Mexico}
\affiliation{Nikhef, Science Park, 1098 XG Amsterdam, the Netherlands}
\affiliation{Radboud University Nijmegen, 6525 AJ Nijmegen, the Netherlands}
\affiliation{Joint Institute for Nuclear Research, Dubna 141980, Russia}
\affiliation{Institute for Theoretical and Experimental Physics, Moscow 117259, Russia}
\affiliation{Moscow State University, Moscow 119991, Russia}
\affiliation{Institute for High Energy Physics, Protvino, Moscow region 142281, Russia}
\affiliation{Petersburg Nuclear Physics Institute, St. Petersburg 188300, Russia}
\affiliation{Instituci\'{o} Catalana de Recerca i Estudis Avan\c{c}ats (ICREA) and Institut de F\'{i}sica d'Altes Energies (IFAE), 08193 Bellaterra (Barcelona), Spain}
\affiliation{Uppsala University, 751 05 Uppsala, Sweden}
\affiliation{Taras Shevchenko National University of Kyiv, Kiev, 01601, Ukraine}
\affiliation{Lancaster University, Lancaster LA1 4YB, United Kingdom}
\affiliation{Imperial College London, London SW7 2AZ, United Kingdom}
\affiliation{The University of Manchester, Manchester M13 9PL, United Kingdom}
\affiliation{University of Arizona, Tucson, Arizona 85721, USA}
\affiliation{University of California Riverside, Riverside, California 92521, USA}
\affiliation{Florida State University, Tallahassee, Florida 32306, USA}
\affiliation{Fermi National Accelerator Laboratory, Batavia, Illinois 60510, USA}
\affiliation{University of Illinois at Chicago, Chicago, Illinois 60607, USA}
\affiliation{Northern Illinois University, DeKalb, Illinois 60115, USA}
\affiliation{Northwestern University, Evanston, Illinois 60208, USA}
\affiliation{Indiana University, Bloomington, Indiana 47405, USA}
\affiliation{Purdue University Calumet, Hammond, Indiana 46323, USA}
\affiliation{University of Notre Dame, Notre Dame, Indiana 46556, USA}
\affiliation{Iowa State University, Ames, Iowa 50011, USA}
\affiliation{University of Kansas, Lawrence, Kansas 66045, USA}
\affiliation{Louisiana Tech University, Ruston, Louisiana 71272, USA}
\affiliation{Northeastern University, Boston, Massachusetts 02115, USA}
\affiliation{University of Michigan, Ann Arbor, Michigan 48109, USA}
\affiliation{Michigan State University, East Lansing, Michigan 48824, USA}
\affiliation{University of Mississippi, University, Mississippi 38677, USA}
\affiliation{University of Nebraska, Lincoln, Nebraska 68588, USA}
\affiliation{Rutgers University, Piscataway, New Jersey 08855, USA}
\affiliation{Princeton University, Princeton, New Jersey 08544, USA}
\affiliation{State University of New York, Buffalo, New York 14260, USA}
\affiliation{University of Rochester, Rochester, New York 14627, USA}
\affiliation{State University of New York, Stony Brook, New York 11794, USA}
\affiliation{Brookhaven National Laboratory, Upton, New York 11973, USA}
\affiliation{Langston University, Langston, Oklahoma 73050, USA}
\affiliation{University of Oklahoma, Norman, Oklahoma 73019, USA}
\affiliation{Oklahoma State University, Stillwater, Oklahoma 74078, USA}
\affiliation{Oregon State University, Corvallis, Oregon 97331, USA}
\affiliation{Brown University, Providence, Rhode Island 02912, USA}
\affiliation{University of Texas, Arlington, Texas 76019, USA}
\affiliation{Southern Methodist University, Dallas, Texas 75275, USA}
\affiliation{Rice University, Houston, Texas 77005, USA}
\affiliation{University of Virginia, Charlottesville, Virginia 22904, USA}
\affiliation{University of Washington, Seattle, Washington 98195, USA}
\author{V.M.~Abazov} \affiliation{Joint Institute for Nuclear Research, Dubna 141980, Russia}
\author{B.~Abbott} \affiliation{University of Oklahoma, Norman, Oklahoma 73019, USA}
\author{B.S.~Acharya} \affiliation{Tata Institute of Fundamental Research, Mumbai-400 005, India}
\author{M.~Adams} \affiliation{University of Illinois at Chicago, Chicago, Illinois 60607, USA}
\author{T.~Adams} \affiliation{Florida State University, Tallahassee, Florida 32306, USA}
\author{J.P.~Agnew} \affiliation{The University of Manchester, Manchester M13 9PL, United Kingdom}
\author{G.D.~Alexeev} \affiliation{Joint Institute for Nuclear Research, Dubna 141980, Russia}
\author{G.~Alkhazov} \affiliation{Petersburg Nuclear Physics Institute, St. Petersburg 188300, Russia}
\author{A.~Alton$^{a}$} \affiliation{University of Michigan, Ann Arbor, Michigan 48109, USA}
\author{A.~Askew} \affiliation{Florida State University, Tallahassee, Florida 32306, USA}
% ...
%
%
\collaboration{The D0 Collaboration\footnote{with visitors from
$^{a}$Augustana College, Sioux Falls, SD 57197, USA,
$^{b}$The University of Liverpool, Liverpool L69 3BX, UK,
$^{c}$Deutshes Elektronen-Synchrotron (DESY), Notkestrasse 85, Germany,
$^{d}$CONACyT, M-03940 Mexico City, Mexico,
$^{e}$SLAC, Menlo Park, CA 94025, USA,
$^{f}$University College London, London WC1E 6BT, UK,
$^{g}$Centro de Investigacion en Computacion - IPN, CP 07738 Mexico City, Mexico,
$^{h}$Universidade Estadual Paulista, S\~ao Paulo, SP 01140, Brazil,
$^{i}$Karlsruher Institut f\"ur Technologie (KIT) - Steinbuch Centre for Computing (SCC),
D-76128 Karlsruhe, Germany,
$^{j}$Office of Science, U.S. Department of Energy, Washington, D.C. 20585, USA,
$^{k}$American Association for the Advancement of Science, Washington, D.C. 20005, USA,
$^{l}$Kiev Institute for Nuclear Research (KINR), Kyiv 03680, Ukraine,
$^{m}$University of Maryland, College Park, MD 20742, USA,
$^{n}$European Orgnaization for Nuclear Research (CERN), CH-1211 Geneva, Switzerland,
$^{o}$Purdue University, West Lafayette, IN 47907, USA,
and
$^{p}$Institute of Physics, Belgrade, Belgrade, Serbia.
}} \noaffiliation
\vskip 0.25cm

\date{Oct. 11, 2017}
\begin{abstract}
We present a measurement of the effective weak mixing angle parameter $\effstw$, 
in $\DY$ events at a center of mass energy of 1.96 TeV, collected by the D0 detector at the Fermilab Tevatron 
Collider and corresponding 
to 8.6 fb$^{-1}$ of integrated luminosity.     
The measured value of $\effstw[\mu\mu]=0.23016 \pm 0.00064$ is further combined with the result 
from the D0 measurement in $p\bar{p}\rightarrow Z/\gamma^{*}\rightarrow e^{+} e^{-}$ events, resulting in $\effstw [\text{comb.}]=0.23095 \pm 0.00040$. 
This combined result is the most precise measurement from a single experiment at a hadron collider and is the most 
precise determination using the coupling of the $Z/\gamma^*$ to light quarks.
\end{abstract}
\pacs{12.15.-y, 12.15.Mm, 13.85.Qk, 14.70.Hp}
\maketitle

The weak mixing angle $\theta_W$ is a fundamental parameter of the standard model (SM). 
It governs the mechanism of spontaneous symmetry breaking of $SU(2)\times U(1)$ in which 
the original vector boson fields $W$ and $B_0$ are transformed to the physical $W^{\pm}$, $Z$ 
and $\gamma$ states. At tree level and in all orders of the on-shell renormalization scheme, 
the weak mixing angle also relates the $W$ and $Z$ boson masses by $\sin^2\theta_W = 1 - M^2_W / M^2_Z$. 
To include higher order electroweak radiative corrections and allow comparison with 
experimental measurements, the effective weak mixing angle can be defined~\cite{LEP-SLD} in terms 
of the relative strengths of the axial vector and vector couplings, $g^f_A$ and $g^f_V$, of the $Z$ 
boson to fermions, $f$:

\begin{equation}
\sin^2\theta^f_{\text{eff}}=\frac{1}{4|Q_f|}\left(1-\frac{g_V^f}{g_A^f}\right)
\end{equation}
where $Q_f$ is the electric charge of the fermions.

It is customary to quote the charged lepton effective weak mixing angle parameter $\effstw$,
determined by measurements of observables around the $Z$ boson mass pole ($M_Z$).
The effective mixing angle was precisely measured by the LEP Collaborations and the SLD Collaboration. 
The combined LEP and SLD result~\cite{LEP-SLD} gives a value of $\effstw = 0.23153 \pm 0.00016$ at 
the energy scale $\mu = M_Z$, while 
the two most precise measurements, 
the value of $0.23221\pm 0.00029$ obtained from the measurement of $b$-quark forward-backward 
asymmetry at LEP, and the value of $0.23098\pm0.00026$ obtained from the measurement of the 
left-right polarization asymmetry at SLD, differ by 3.2 standard deviations. The large difference 
might indicate a possible bias in one of the measurements and in the world average. A precise independent 
determination of the effective weak mixing angle is therefore an important  
test of the SM electroweak breaking mechanism.

At the Tevatron, the weak mixing angle can be measured in the Drell-Yan
process $p\bar{p} \rightarrow Z/\gamma^{*} \rightarrow \ell^+\ell^-$ 
through a forward-backward charge asymmetry, $A_{FB}$, defined by 
$A_{FB} = (N_F - N_B) / (N_F + N_B)$, where $N_F$ and $N_B$ are the numbers 
of forward and backward events. 
Forward ($F$) or backward ($B$) events are defined as those for which $\cos\theta^* > 0$ 
or $\cos\theta^* < 0$, where $\theta^*$ is the angle between the negatively charged lepton 
direction and the incoming proton direction in the Collins-Soper frame~\cite{cs-frame}. 

For the $Z$ to fermion couplings, both $g^f_A = I^f_3$ and $g^f_V = I^f_3 - 2 Q_f \sin^2\theta_W$ exist, 
whereas for the photon to fermion couplings there is only a vector coupling. $I^f_3$ is the third component of the weak isospin 
of the fermion. The parity violation implicit in the forward-backward asymmetry 
arises from the interference between the vector and axial vector couplings. As the main subprocess for 
Drell-Yan production is the quark-antiquark annihilation $q\bar{q}\rightarrow \ell^+\ell^-$, $A_{FB}$ depends upon 
both the couplings to light quarks and the couplings to leptons.
The asymmetry can be measured as a function of the
invariant mass of the dilepton pair.
Since only the vector 
coupling of the $Z$ boson depends on $\sin^2\theta_W$, the information on $\sin^2\theta_W$ comes from 
the asymmetry in the vicinity of the $Z$ boson pole. 
Away from the $Z$ boson mass pole, the asymmetry results from the interference of the axial vector $Z$ coupling and 
vector photon coupling and depends upon the parton distribution functions (PDFs).

Measurements of $\effstw$ corresponding to the full dataset at the Fermilab 
Tevatron Collider were performed by the CDF Collaboration using 
$Z/\gamma^*\rightarrow \mu^+\mu^-$ channel~\cite{cdf_muon} and $Z/\gamma^*\rightarrow e^+e^-$
channel~\cite{cdf_stw_final}, and by the D0 Collaboration in the 
$Z/\gamma^*\rightarrow e^+e^-$ channel~\cite{d0-zee-10fb}. 
The weak mixing angle was also measured at the Large Hadron Collider (LHC) by 
the ATLAS, CMS and LHCb collaborations~\cite{atlas_pub,cms_pub,lhcb_pub}. 
Because the directions of the initial quarks and anti-quarks in the dominant subprocess 
$q\bar{q} \rightarrow Z/\gamma^* \rightarrow \ell^+\ell^-$ process 
are unknown and have to be estimated in $pp$ collisions, 
the precision of the LHC results is not as good as that of the Tevatron even with higher statistics. 

This article reports a measurement of the effective weak mixing angle from the $A_{FB}$ distribution as a function 
of the dimuon invariant mass using 
8.6~fb$^{-1}$ of data collected by the D0 detector at the Fermilab Tevatron Collider using the $Z/\gamma^*\rightarrow \mu^+\mu^-$ 
channel. The $Z/\gamma^* \rightarrow \mu^+\mu^-$ measurement is then combined with 
the D0 $Z/\gamma^* \rightarrow e^+e^-$ measurement~\cite{d0-zee-10fb}. 

\indent The D0 detector comprises a central tracking
system, a calorimeter and a muon system~\cite{d0-detector, d0-det-2, d0-det-3}. The central tracking
system consists of a silicon microstrip tracker and a scintillating fiber tracker,
both located within a 1.9~T superconducting solenoidal magnet and optimized for tracking and vertexing
capabilities for detector pseudorapidities of $|\eta_{\text{det}}|<3$~\cite{d0_coordinate}.
Outside the solenoid, three liquid-argon and uranium calorimeters provide coverage for
$|\eta_{\text{det}}|<3.5$ for electrons. 
The muon system is located outside of the calorimeters, providing coverage for 
$|\eta_\text{det}|<2.0$. It consists of drift chambers and scintillators and  
1.8~T iron toroidal magnets. The solenoid and toroid polarities are reversed every two weeks on average to 
reduce detector-induced asymmetries. 
Muons are identified using information from both the tracking system and the muon system. 
Muon momenta are measured using the tracking system information.

To maximize the event sample, 
data collected with all triggers are used in this analysis. 
Events are required to have at least two muon candidates reconstructed in the tracking system and 
the muon system. Both muon candidates~\cite{muonID_nim} are required to have transverse momentum $p_T > 15$ GeV$/c$ and 
$|\eta|<1.8$ with at least one muon within $|\eta|<1.6$. The two 
muon candidates must be isolated from jets in the event by requiring the sum 
of transverse momenta of tracks in the tracking system or transverse energy in the calorimeter 
within cones surrounding the muon candidate to be small. Muons must have a track in 
the tracking system matched with one in the muon system. To suppress backgrounds, the 
two matched tracks are required to point to the same $p\bar{p}$ interaction vertex and to 
have opposite charges. Events with muons nearly back-to-back are removed to reduce cosmic 
ray background. Events are further required to have a reconstructed dimuon 
invariant mass $74 < M_{\mu\mu} < 110$ GeV$/c^2$. The number of
events satisfying these requirements is 481,239. 

The Monte Carlo (MC) Drell-Yan $Z/\gamma^{*} \rightarrow \mu^+\mu^-$ sample is generated
using leading-order {\sc pythia}~\cite{pythia} with the NNPDF3.0~\cite{nnpdf3} PDFs, followed 
by a {\sc geant}-based simulation~\cite{geant} of the D0 detector. 
Events from randomly selected beam crossings with the same instantaneous luminosity profile as data 
are overlaid on the simulated events to model detector noise and contributions from the presence of 
additional $p\bar{p}$ interactions.
The {\sc pythia} MC samples are used to study the detector's geometric 
acceptance and the momentum scale and resolution of muons. Separate MC samples are generated 
for the four different polarity combinations of the solenoid and toroid magnetic fields. 

The effective weak mixing angle, which is extracted from $A_{FB}$ as a function of $M_{\mu\mu}$, 
depends strongly on the dimuon mass calibration. Therefore, it 
is critical to have a precise muon momentum measurement and a consistent measured 
mean value of $M_{\mu\mu}$ for all $\eta$, and each muon charge sign $q$ and solenoid polarity $S$. 
The D0 muon momentum calibration and resolution smearing procedure~\cite{muonID_nim} is 
applied to the MC simulation, so as to give agreement of the overall width and peak value of the $M_{\mu\mu}$ distribution 
with data. However, the muon momentum measurement, 
especially the scale of the reconstructed muon momentum, 
still depends on the charge and $\eta$ of the muons due to imperfect 
alignment of the detector~\cite{misalignment}.
Such dependence 
would translate into a large systematic uncertainty on the $A_{FB}$ measurement. 
To reduce this dependence, an additional  
correction, $\alpha(q, \eta, S)$, to the muon momentum is applied to the data and MC separately. 
This factor is determined by requiring the mean of the 
$M_{\mu\mu}$ distribution over the full mass range in each $(q, \eta, S)$ region to be consistent with the corresponding nominal 
value obtained from a generator-level MC sample after applying the same kinematic and acceptance 
cuts as those applied to the data. After the calibration, the mean values of $M_{\mu\mu}$ 
in data and MC are consistent to within statistical fluctuations. 
The additional calibration, together with the D0 muon calibration 
and resolution smearing procedure~\cite{muonID_nim}, reduces not only the $q$-$\eta$-$S$ dependence, but also 
the potential effect from an imperfect modeling on the final state radiation in the {\sc pythia} generator.
The residual difference between data and MC 
$M_{\mu\mu}$ mean values is propagated to the uncertainty of the weak mixing angle measurement.

Additional corrections and reweightings are applied to the MC simulation to 
improve the agreement with data. 
The ratio between the MC and data efficiencies for the muon identification 
is measured using the tag-and-probe 
method~\cite{muonID_nim} and applied to the MC distributions as a function of muon $\eta$. 
The simulation is further corrected for higher-order effects not included in {\sc pythia}
by reweighting the MC events at the generator level in two dimensions
($p_{T}$ and rapidity $y$ of the $Z$ boson)
to match {\sc resbos}~\cite{resbos} predictions. In addition, next-to-next-to-leading order QCD
corrections are applied as a function of $Z$ boson mass~\cite{resbos, NNLO_QCD}. 

The sign of the track matched to the muon is used to determine the charge of the muon and 
to classify the event as forward or backward. The charge misidentification rate measured in the 
data is smaller than $0.4\%$. Since the opposite charge sign requirement is applied in 
the event selection, the probability of both muons charges to be misidentified, 
thus transforming a forward event into a backward event or vice versa, is negligibly small.

Background is suppressed by the strict requirements on the muon tracks. The main 
remaining contribution is from multijet events, in which jets are misidentified as muons, and is estimated from 
data by selecting events with reversed muon isolation cuts 
in order to study the shape of the mass distribution of multijet events. 
The normalization of the multijet background is assumed to be same as that of the selected 
same sign events after 
correcting for the presence of the misidentified signal events and the additional background contributions 
described below.
The $W+$jets background is generated using {\sc alpgen}~\cite{alpgen} interfaced to 
{\sc pythia} for showering and hadronization. The 
$Z/\gamma^*\rightarrow \tau\tau$, di-boson and $t\bar{t}$ backgrounds, are estimated 
using {\sc pythia}. 
In the dimuon mass range used for the effective weak mixing angle measurement, the multijet 
background is $0.68\% \pm 0.68\%$. An $100\%$ uncertainty is used to safely cover the bias 
due to corrections for the misidentified signal events.
The sum of the $W+$jets,  $Z/\gamma^*\rightarrow \tau\tau$, di-boson ($WW$ and $WZ$) 
and $t\bar{t}$ background is $0.20\% \pm 0.05\%$, where the uncertainty is mainly from 
cross sections of the physics backgrounds. 

\begin{figure}[!hbt]
\begin{center}
 \epsfig{scale=0.45, file=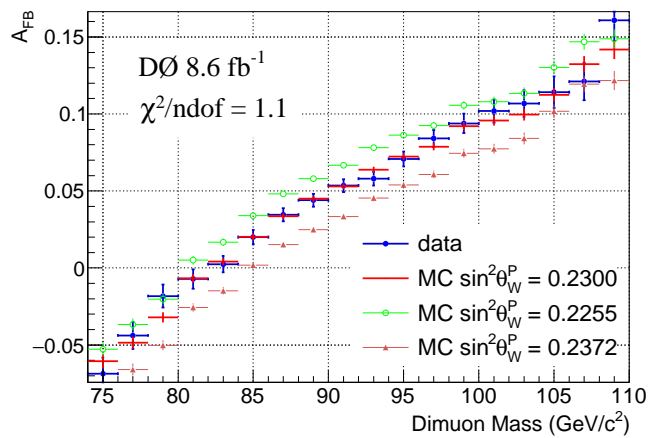}
 \caption{\small (color online). Comparison between the $A_{FB}$ distributions in
 the background-subtracted data and the MC 
 with different $\sin^2\theta^\text{p}_W$ values in the {\sc pythia} generator. 
 The $\chi^2$ corresponds to the MC with the best fit value of $\sin^2\theta^\text{p}_W$. The uncertainties are statistical only.}
 \label{fig:rawAFB}
\end{center}
\end{figure}

The effective weak mixing angle is extracted from the background-subtracted $A_{FB}$ spectrum 
by comparing the data to simulated $A_{FB}$ templates corresponding to different input values 
of the weak mixing angle. The effective weak mixing angle parameter, here denoted as $\sin^2\theta^\text{p}_W$, 
corresponds to the input parameter in the calculation from the leading order {\sc pythia} generator. 
Higher order corrections are used to convert $\sin^2\theta^\text{p}_W$ to $\effstw$~\cite{TevatronCombine}.
The templates are obtained by reweighting the two-dimensional distribution of the 
$Z$ boson mass and $\cos\theta^*$ at the generator level 
to different $\sin^2\theta^\text{p}_W$ {\sc pythia} predictions. 
The background-subtracted $A_{FB}$
distribution and {\sc pythia} predictions are
shown in Fig.~\ref{fig:rawAFB}.

The uncertainties on the fitted $\sin^2\theta^\text{p}_W$, listed in Table~\ref{tab:result_muon}, are dominated by the limited size 
of the data sample. 
The systematic uncertainties due to muon momentum calibration and resolution smearing, 
the estimation of the backgrounds and the efficiency scale factors are themselves also dominated 
by the limited data samples. 
The PDF uncertainty is obtained as the standard deviation of the distribution of $\sin^2\theta^\text{p}_W$ values 
given by each of the equal-weighted PDF sets from NNPDF3.0~\cite{nnpdf3}. The best fit is 

\begin{eqnarray*}
 \sin^2\theta^\text{p}_W &=& 0.22994 \pm 0.00059\thinspace (\text{stat.}) \pm \\
 &&0.00005\thinspace (\text{syst.})\pm 0.00024\thinspace (\text{PDF}).
\end{eqnarray*}

\begin{table}
\begin{center}
\begin{tabular}{l|c}
\hline \hline
    $\sin^2\theta^\text{p}_W$  &   0.22994 \\ \hline
   Statistical uncertainty &  0.00059 \\
\hline
   Systematic   &  \\
   ~~Momentum calibration      &  0.00002 \\
   ~~Momentum smearing      &  0.00004 \\
   ~~Background       &  0.00003 \\
   ~~Efficiencies      &  0.00001 \\
   ~~Total systematic  & 0.00005 \\
\hline
   PDF  & 0.00024 \\
\hline
   Total   & 0.00064 \\
\hline \hline
\end{tabular}
\caption{\small Measured $\sin^2\theta^\text{p}_W$ value and corresponding uncertainties. 
All uncertainties are symmetric. 
Higher order corrections are not included.}
\label{tab:result_muon}
\end{center}
\end{table}

Higher order corrections from several sources are introduced. 
The measured $\sin^2\theta^\text{p}_W$ is an average over the leptonic, $u$-quark and $d$-quark 
effective couplings~\cite{d0-zee-10fb}. The mass-scale dependence and complex valued calculation of 
the weak corrections and the fermion-loop correction to the photon propagator, 
are not considered in the {\sc pythia} generator~\cite{TevatronCombine}. 
To obtain the leptonic effective weak mixing angle parameter, we shift the central value 
by $+0.00022$, of which $+0.00008$ is for the $u/d$-quark correction and 
$+0.00014$ is for the complex valued calculation and mass-scale dependence correction~\cite{TevatronCombine}.
An additional systematic uncertainty of 0.00004 is further introduced~\cite{TevatronCombine}.
After higher order corrections applied, we get $\effstw[\mu\mu] = 0.23016 \pm 0.00064$. 

The D0 $e^+e^-$ measurement~\cite{d0-zee-10fb} and the $\mu^+\mu^-$ measurement 
presented here are used as inputs to a D0 combination result for $\effstw$. The 
$e^+e^-$ measurement in Ref.~\cite{d0-zee-10fb} has been modified 
for consistency to incorporate the use of additional higher order corrections and the NNPDF3.0 PDFs 
employed in this letter and in the CDF measurement~\cite{cdf_stw_final}.
The corrected value is 
$\effstw[ee]=0.23137\pm 0.00047$~\cite{TevatronCombine}. The D0 
$e^+e^-$ and $\mu^+\mu^-$ measurements agree to within 1.4 standard deviations. 

The central values and systematic uncertainties of the $e^+e^-$ and $\mu^+\mu^-$ channels 
are combined using the inverse of the squares of the statistical uncertainties as weights. 
The systematic uncertainties are treated as uncorrelated, except the higher order correction uncertainty 
which is treated as $100\%$ correlated.
However, the total combined uncertainty in practice does not depend on whether 
the systematic uncertainties of the input measurements are taken to be correlated or uncorrelated, because both 
measurements are dominated by statistical uncertainties. 
The correlation of the acceptances between $e^+e^-$
and $\mu^+\mu^-$ channels cannot be ignored in treating the PDF uncertainty. 
Instead of estimating a correlation matrix 
between $\effstw$ results for these two channels, a combined PDF uncertainty is estimated by first estimating 
the PDF uncertainty 
on the average of values for the $e^+e^-$ and $\mu^+\mu^-$ 
channels, and then scaling that uncertainty using the linear relation between $A_{FB}$ and 
$\sin^2\theta^\text{p}_W$ calculated using MC.

\begin{figure}[!hbt]
\begin{center}
 \epsfig{scale=0.45, file=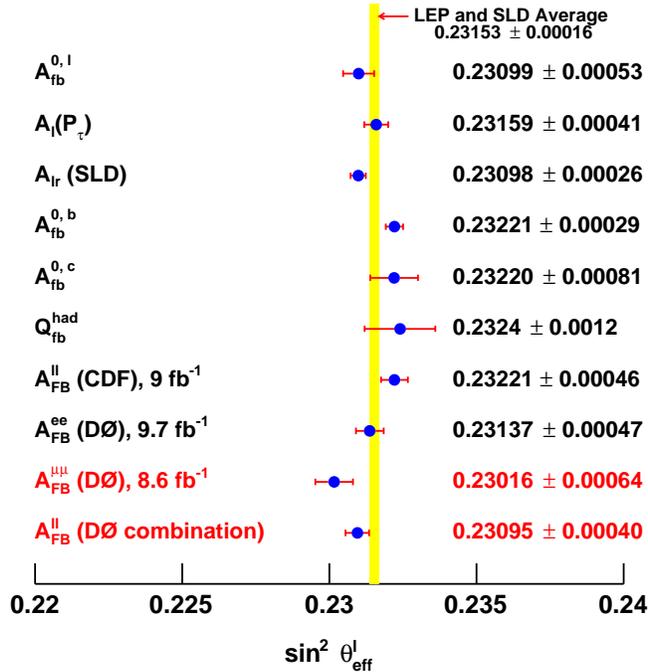}
\caption{\small (color online). Comparison of $\effstw(M_Z)$ measured by D0 with
results from other experiments.
The average of measurements from the LEP and SLD Collaborations~\cite{LEP-SLD} is 
also shown.}
\label{fig:stw_combined}
\end{center}
\end{figure}

The combination is:

\begin{eqnarray*}
 \effstw[\text{comb.}] &=& 0.23095 \pm 0.00035\thinspace (\text{stat.}) \pm \\
 &&0.00007\thinspace (\text{syst.})\pm 0.00019\thinspace (\text{PDF}).
\end{eqnarray*}
Table~\ref{tab:result_comb} summarizes the inputs and the results of the 
combination of the $e^+e^-$ and $\mu^+\mu^-$ measurements. 
The measured $\effstw$ from D0 and other experiments are compared to the 
LEP and SLD average in Fig.~\ref{fig:stw_combined}. The 
D0 combination has an uncertainty close to the precision of the world's best
measurements performed by the LEP and SLD Collaborations.

\begin{table}
\begin{center}
\begin{tabular}{l|c|c|c}
\hline \hline
      &   $e^+e^-$ channel & $\mu^+\mu^-$ channel & Combined \\
\hline
   $\effstw$ & 0.23137 & 0.23016 & 0.23095 \\
\hline
   Statistical & 0.00043 & 0.00059 & 0.00035 \\
\hline
   Systematic & 0.00009 & 0.00006 & 0.00007 \\
\hline
   PDF & 0.00017 & 0.00024 & 0.00019 \\
\hline
   Total & 0.00047 & 0.00064 & 0.00040\\
\hline \hline
\end{tabular}
\caption{\small Combined measurement of $\effstw$ and breakdown of its uncertainties, 
together with the corresponding input values.
All uncertainties are symmetric.}
\label{tab:result_comb}
\end{center}
\end{table}

The measurement of $\effstw$ can be used to determine the on-shell value 
of $\stw$ and the mass of the $W$ boson, $M_W$. The relationships between the on-shell $\stw$, 
$\effstw$ and $M_W$ are: 

\begin{eqnarray*}
  \effstw &=& \text{Re}[\kappa_e(M^2_Z)] \sin^2\theta_W, \\
  \sin^2\theta_W &=& 1 - \frac{M^2_W}{M^2_Z},
\end{eqnarray*}
  where $\text{Re}[\kappa_e(M^2_Z)]$ is a form factor 
  whose value at the $Z$ mass pole is 1.037~\cite{formfactor}. 
  The on-shell $\sin^2\theta_W$ and the indirect determination of $M_W$ based on 
  the D0 combined measurement of $\effstw$ are

\begin{eqnarray*}
  \sin^2\theta_W &=& 0.22269 \pm 0.00040 \\
  M_W &=& 80396 \pm 21 \text{ MeV}/c^2.
\end{eqnarray*}

In conclusion, we have measured the effective weak mixing angle parameter from the 
forward-backward charge asymmetry $A_{FB}$ distribution in the process 
$p\bar{p}\rightarrow Z/\gamma^* \rightarrow \mu^+\mu^-$ at the Fermilab Tevatron Collider. 
The primary systematic uncertainty arising from muon momentum calibration 
is reduced by introducing a charge-$\eta$-solenoid-dependent calibration. 
The final result using 8.6~fb$^{-1}$ of D0 Run II data is $\effstw[\mu\mu]=0.23016 \pm 0.00064$, which 
is at the level of the best single channel precision from hadron collider experiments. The D0 combination 
of the $e^+e^-$ and $\mu^+\mu^-$ measurements is $\effstw[\text{comb.}]=0.23095 \pm 0.00040$, which 
is the most precise single experiment measurement at hadron colliders and is the most 
precise result based on the coupling of light quarks to the $Z$ boson.

\section{Acknowledgements}
This document was prepared by the D0 collaboration using the resources of the Fermi National Accelerator Laboratory (Fermilab),
 a U.S. Department of Energy, Office of Science, HEP User Facility. Fermilab is managed by Fermi Research Alliance, LLC (FRA),
acting under Contract No. DE-AC02-07CH11359.  

We thank the staffs at Fermilab and collaborating institutions,
and acknowledge support from the
Department of Energy and National Science Foundation (United States of America);
Alternative Energies and Atomic Energy Commission and
National Center for Scientific Research/National Institute of Nuclear and Particle Physics  (France);
Ministry of Education and Science of the Russian Federation, 
National Research Center ``Kurchatov Institute" of the Russian Federation, and 
Russian Foundation for Basic Research  (Russia);
National Council for the Development of Science and Technology and
Carlos Chagas Filho Foundation for the Support of Research in the State of Rio de Janeiro (Brazil);
Department of Atomic Energy and Department of Science and Technology (India);
Administrative Department of Science, Technology and Innovation (Colombia);
National Council of Science and Technology (Mexico);
National Research Foundation of Korea (Korea);
Foundation for Fundamental Research on Matter (The Netherlands);
Science and Technology Facilities Council and The Royal Society (United Kingdom);
Ministry of Education, Youth and Sports (Czech Republic);
Bundesministerium f\"{u}r Bildung und Forschung (Federal Ministry of Education and Research) and 
Deutsche Forschungsgemeinschaft (German Research Foundation) (Germany);
Science Foundation Ireland (Ireland);
Swedish Research Council (Sweden);
China Academy of Sciences and National Natural Science Foundation of China (China);
and
Ministry of Education and Science of Ukraine (Ukraine).

We thank Dr. W. Sakumoto for his help in assuring that 
the CDF and D0 collaborations used a similar phenomenological framework for these measurements.


\begin{thebibliography}{99}
\bibitem{d0_coordinate}
D0 uses a cylindrical coordinate system with the $z$ axis
along the beam axis in the proton direction. Angles $\theta$ and $\phi$ are the polar and
azimuthal angles, respectively. Pseudorapidity is defined as
$\eta=-\ln [\tan(\theta/2)]$ where $\theta$ is measured with respect
to the interaction vertex. In the massless limit, $\eta$ is equivalent to
the rapidity $y=(1/2) \ln[(E+p_z)/(E-p_z)]$,
and $\eta_{\text{det}}$ is the pseudorapidity measured with respect to the
center of the detector.
\end{thebibliography}
\end{document}